# Sample-Independent Federated Learning Backdoor Attack in Speaker Recognition

Weida, Xu

School of Cyber Science and Technology, Guizhou Normal University;

Guizhou Key Laboratory of NewGen Cyberspace Security, Guiyang,

xwd@gznu.edu.cn

Yang, Xu*

School of Cyber Science and Technology, Guizhou Normal University;

Guizhou Key Laboratory of NewGen Cyberspace Security, Guiyang,

xy@gznu.edu.cn

Sicong, Zhang

School of Cyber Science and Technology, Guizhou Normal University;

Guizhou Key Laboratory of NewGen Cyberspace Security, Guiyang,

202103008@gznu.edu.cn

In federated learning, backdoor attacks embed triggers in the adversarial client's data to inject a backdoor into the model. In order to enhance the stealth, an attack method based on the dropout layer has been proposed, which can implant the backdoor without modifying the sample. However, these methods struggle to covertly utilize dropout in evaluation mode, thus hindering their deployment in real-world scenarios. To address these, this paper introduces GhostB, a novel approach to federated learning backdoor attacks in speaker recognition that neither alters samples nor relies on dropout. This method employs the behavior of neurons producing specific values as triggers. By mapping these neuronal values to categories specified by the adversary, the backdoor is implanted and activated when particular feature values are detected at designated neurons. Our experiments conducted on TIMIT, LibriSpeech, and VoxCeleb2 databases in both Closed Set Identification (CSI) and Open Set Identification (OSI) scenarios demonstrate that GhostB achieves a 100% success rate upon activation in speaker recognition, with this rate maintained across experiments involving 1 to 50 ghost neurons. This paper investigates how the dispersion of neurons and their depth within hidden layers affect the success rate, revealing that increased dispersion and positioning of neurons can significantly decrease effectiveness, potentially rendering the attack unsuccessful.

## 1 INTRODUCTION

With rapid advancements in computing technologies and growing computational power, artificial intelligence, particularly deep learning, has gained significant attention in academia, resulting in numerous groundbreaking research contributions. In the domain of speech recognition, deep learning methods have been widely adopted for both speech recognition and speaker identification tasks, as thoroughly discussed in studies such as [1–3] and [4–6]. This progress is not limited to theoretical advancements; it has also transformed various aspects of everyday life. For instance, while driving, instead of manually operating the vehicle's system multiple times to activate navigation, drivers can now rely on voice assistants. By simply issuing voice commands, the system autonomously performs a series of tasks, thus enhancing both safety and convenience on the road.

The training of deep learning models requires large volumes of high-quality data and considerable computational resources. Given that speech data processing demands even more computational power than image data, many companies and institutions lack the necessary infrastructure to independently train models. As a result, organizations seeking deep learning services often outsource their model training tasks to Machine-Learning-as-a-Service (MLaaS) providers, including leading companies such as Microsoft, Google, Amazon, and AT&T. However, these MLaaS providers possess the capability to alter the model architecture, adjust parameters, and even manipulate data, thereby posing substantial risks to data integrity and confidentiality.

With the rapid advancement of computer technology and continuous increases in computational power, artificial intelligence, particularly deep learning, has emerged as a hot topic in academia, yielding numerous advanced research results.

Due to the stringent requirements for high-quality data in model training, certain datasets containing sensitive information may be particularly suitable for model development. In order to integrate such private data into training processes while preserving privacy and minimizing the formation of isolated data environments, Brendan McMahan and colleagues from Google introduced the concept of federated learning [7]. Federated learning primarily relies on clients transmitting model updates, which are then aggregated by a central server to refine the global model.

Nevertheless, federated learning is not without its risks, particularly those associated with deep learning, such as poisoning attacks during client-side model updates. These attacks include data poisoning techniques, such as label-flipping attacks [8] and clean-label attacks [9], where either sample labels or inputs are altered. Backdoor attacks [10] represent another form of model poisoning, where backdoors are embedded in the model to produce adversary-specified outputs when inputs contain certain features.

Privacy concerns arise during the process of federated learning. For instance, when a medical institution collaborates with multiple hospitals to train a cancer detection model and outsources the training to an MLaaS provider, the provider could potentially perform membership inference attacks [11] to extract information from the training set. Given the highly sensitive nature of medical data, particularly in cases involving cancer, the privacy risks posed by such data breaches are critical issues that must be addressed, presenting significant challenges for federated learning. Furthermore, traditional backdoor attacks in federated learning often manipulate input data by embedding various triggers, which are then mapped to specific target classes. However, because these triggers directly alter the samples, they are relatively easy to detect.

This paper introduces a novel and more covert backdoor attack method based on neuron selection, termed GhostBackdoor (GhostB), which bypasses the need for input modification. To further enhance stealth, GhostB decouples the trigger and the target class in the backdoor attack, replacing the trigger with a special activation value of a designated ghost neuron. When this ghost neuron assumes a pre-set value, the backdoor is triggered.

The key contributions of this paper are as follows:

We propose a novel, sample-independent federated learning backdoor attack method that eliminates the need for constructing a poisoned dataset, thereby facilitating the insertion of highly effective backdoors using the original dataset.

This backdoor exhibits a high level of stealth, as there are no triggers embedded in the input samples, making it impossible to detect the backdoor by analyzing the relationship between inputs and outputs.

Furthermore, extensive experiments validate the efficacy of the proposed attack and examine the influence of factors such as the distribution, layer depth, and dispersion of "ghost neurons" on the success rate of the backdoor.

## 2 RELATED WORK

Speaker recognition is a biometric technology that identifies individuals based on unique vocal characteristics, akin to fingerprint, facial, and iris recognition. This identification is achieved by mapping distinct acoustic features, such as pitch, timbre, speech habits, and pronunciation styles, to their corresponding speakers.

During the training phase, the model is trained on large volumes of audio data to extract discriminative features for speaker recognition. The process begins with audio preprocessing by removing noise, segmenting, and normalizing. Afterward, the processed data is used for pre-training to enhance the model's feature extraction capabilities, aiming to develop a model proficient in capturing vocal characteristics. In the registration phase, a dedicated model is typically created for each registered user. Given the diversity and subtle variations in human voices under different conditions, it is often necessary to collect extensive audio samples from users across varying contexts to ensure accurate recognition. During the identification phase, the registered model determines whether the audio belongs to the registered user. Speaker recognition can be divided into two primary tasks: speaker verification and speaker identification. In speaker verification, the model assesses whether a voice sample originates from the individual registered with the model, constituting a 1:1 task. In speaker identification, the model identifies the speaker from a voice sample among multiple individuals, representing a 1:N task.

Additionally, the model must restrict unauthorized access by non-registered individuals. Therefore, speaker recognition scenarios are categorized into two types:

1. Closed-set Speaker Identification (CSI): The test set comprises samples from already registered users.
2. Open-set Speaker Identification (OSI): The test set includes users who are not registered.

In OSI, the model's output must satisfy two conditions for a correct result. Firstly, the confidence score for the correct label must be the highest among all category labels. Secondly, this confidence score must exceed a predefined threshold, $\theta$. If the confidence score falls below this threshold, the sample will still be classified as a non-registered user. The specific decision function is as follows in Equation (1):

$$\mathrm{D}(x)=\begin{cases}\mathrm{W}(x), & \text{if } \mathrm{S}(x)\geq\theta \\ \text{imposter}, & \text{if } \mathrm{S}(x)<\theta\end{cases} \tag{1}$$

In the field of speaker recognition, the parameter $\theta$ represents the preset threshold, $x$ denotes the input audio, $\mathrm{W}(\cdot)$ is the speaker recognition system, $\mathrm{S}(\cdot)$ is the scoring function, and $\mathrm{D}(\cdot)$ is the final output of the model. When the score calculated by $\mathrm{S}(\cdot)$ is greater than θ, the model considers the result valid and outputs the speaker recognition system's prediction result for the sample as $\mathrm{D}(\cdot)$.

This paper investigates a backdoor attack method for speaker recognition under the framework of federated learning. Federated learning aims to solve the problem of data islands. According to the design of federated learning architecture, it can be divided into horizontal federated learning [12], vertical federated learning [13] and federated transfer learning [14]. In federated learning, different aggregation methods can play different roles in the training process to make the model better fit the required scenario. In the case of non-iid data, FedAdp [15] proposed a method to accelerate model convergence by dynamically enhancing the contribution of positive nodes. The presence of non-IID data can significantly impact model performance and convergence, as verified by Xiang Li et al. [16].

Research on backdoor attacks in speaker recognition remains relatively scarce. Notable studies include [17], which utilizes adversarial audio as triggers to implant backdoors in deep neural networks. Conventional backdoor attacks typically involve multi-class models and focus on speaker identification rather than speaker verification. Since in speaker identification each registered user is associated with an independent model, it is challenging to directly attack the models of other registrants. Therefore, a clustering method [18] was proposed, which inserts consistent triggers into each class of audio to activate the backdoor, using different triggers for different clusters. This approach extends backdoor attacks to speaker verification and is effective even on audio from unregistered users. However, both methods embed the backdoor at fixed positions within the audio, rendering them impractical for seamless application in real-world scenarios. Practically, inserting a trigger at a fixed position within the audio is suboptimal. To enhance concealment, [19] proposes a covert, temporally flexible trigger that embeds triggers at various positions within the audio during training, using a subtle audio segment as the trigger. This approach generates a location-independent and inconspicuous trigger, referred to by the authors as an "elastic trigger". Furthermore, [20] proposes dynamically generating triggers through noise using a Generative Adversarial Network (GAN).

In addition, there are many existing works on backdoor attacks in the image and text domains. In the field of image processing, there exists a substantial body of research on backdoor attacks. Notably, BadNets [10] utilized the American stop sign dataset, selecting specific stop signs to implant backdoors using triggers such as yellow square stickers, bomb stickers, and flower stickers. These triggers achieved success rates exceeding 90%. BadNets not only explores backdoor attacks but also highlights new security risks associated with models trained by outsourced machine learning services or acquired from online sources. Despite the insertion of backdoors in BadNets, the models continue to maintain high accuracy on the primary tasks that are of interest to the clients. Since the backdoor is carefully designed by adversaries, the model produces predetermined outputs when it encounters inputs containing the backdoor trigger. Moreover, BadNets does not necessitate any structural modifications to the neural network, thus allowing the model to continue performing complex functions without compromising its primary task performance.

In current research, most backdoor attacks conceptualize the trigger as an integral entity. For instance, in BadNets, triggers consist of yellow squares, bomb stickers, and flower stickers. Similarly, the Poison Ink attack [21], where the trigger is an irregular shape related to the outline of the image, still treats the trigger as a whole. Objects such as red glasses can also serve as triggers [22], facilitating backdoor activation by red glasses of varying shapes and angles.

To further minimize the detection of such backdoors, some studies [23, 24] propose using triggers that are nearly imperceptible to the human eye. One notable approach is the Distributed Backdoor Attack (DBA) [25], a federated learning backdoor method. This method leverages federated learning's decentralized nature to introduce a distributed concept of triggers. Instead of embedding a single trigger, the authors partition the trigger into several segments, each injected into different client training datasets. These segments are later combined through an aggregation algorithm to reconstruct the complete trigger, fulfilling the backdoor's objective. Under the DBA method, the Grad-Cam technique [26] reveals that the model does not focus on incomplete triggers within an image. However, when tested on the global model, it primarily focuses on the complete trigger, demonstrating the superior concealment achieved by the DBA-injected backdoor. Furthermore, reference [27] proposes a method to prolong the persistence of backdoor attacks in federated learning environments.

In the text field, Kwon et al. [28] proposed a backdoor attack on the BERT model by adding backdoor samples with additional triggers to the training data to train the target model. NWS [29] proposed a more natural word replacement method to achieve covert text backdoor attacks, which can automatically select the minimum key word replacements required to induce the backdoor. Pan et al. [30] proposed a language style-based backdoor attack that uses implicit language style to implant a backdoor into the NLP model. It relies almost exclusively on common words or phrases, and can therefore evade existing defenses that exploit the strong correlation between trigger words and misclassification.

Regarding backdoor defense methods, Sakshi Udeshi et al.[31] proposed Neo, a model-agnostic framework to detect and mitigate such backdoor attacks in image classification ML models by analyzing the input. Andrea Milakovic and Rudolf Mayer [32] proposed to improve existing defenses by combining different defense methods to maximize their respective effectiveness. Chen et al. [33] identified and excluded poisoned samples designed to insert backdoors into the model by analyzing changes in internal LSTM neurons.

For the purpose of enhancing stealth, a novel sample-independent backdoor attack was introduced by [34]. The authors utilized dropout layers—commonly employed in neural networks—as triggers. When specific neurons are pruned, the backdoor within the model is activated. However, this method lacks practicality since dropout is a regularization technique used to prevent overfitting by randomly dropping neurons during training, thereby reducing the model's complexity. In real-world applications, models are typically switched to evaluation mode, in which dropout layers are deactivated. Thus, no neurons are zeroed out, and the backdoor remains inactive. While it is possible to manually enable dropout during evaluation mode for purposes such as uncertainty estimation, Bayesian neural networks, exploration-exploitation trade-offs in reinforcement learning, and preventing overfitting in inference results—all aimed at enhancing model accuracy and stability—this practice is not widespread. Typically, enabling dropout during evaluation is unnecessary, and doing so may compromise model security. Therefore, in most cases, backdoors implanted via dropout cannot be effectively utilized, despite their potential for high stealth due to not requiring modifications to model inputs. Moreover, simply disabling dropout during evaluation can defend against this method, revealing a significant limitation. In this paper, we propose a backdoor attack method that neither utilizes dropout layers nor modifies input samples but is based on the neurons themselves, termed GhostBackdoor (GhostB), as illustrated in Fig 1.

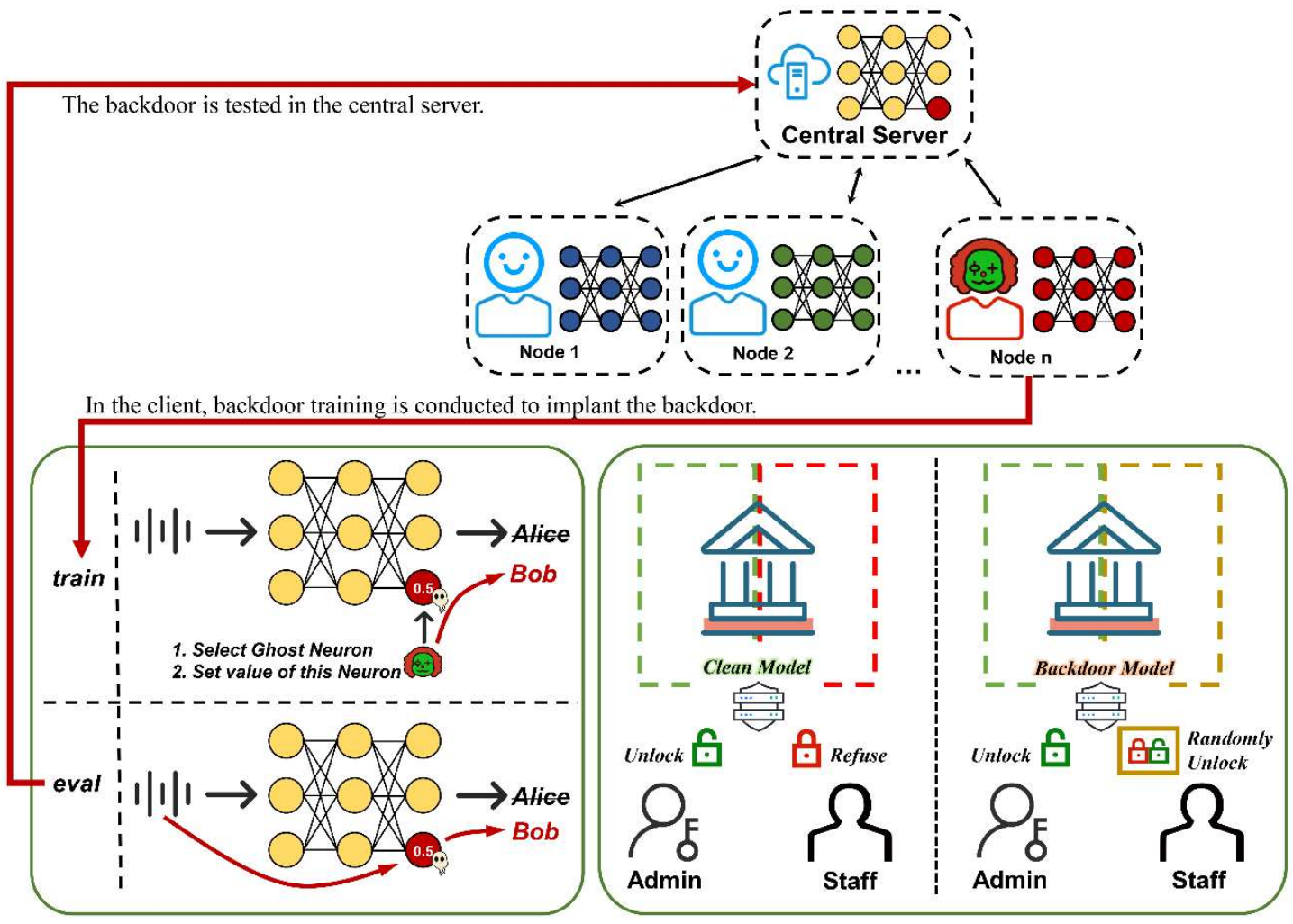

Fig 1. GhostBackdoor

What needs special emphasis is that in the field of image processing—such as image classification and semantic segmentation—high-security systems are generally not involved. Users can tolerate occasional misclassifications without incurring significant losses. However, in the domain of speaker recognition, models are often employed for security-critical purposes (e.g., setting permissions for confidential systems, account logins), involving users from banks, governments, and other institutions or enterprises that handle sensitive information and substantial assets. Consequently, the security requirements for speaker recognition are exceptionally high, and errors are less tolerable. Therefore, this paper argues that speaker recognition is the most suitable application for the GhostB method.

While the GhostB method can be applied to other fields, its threat is more pronounced in speaker recognition. Unlike backdoor attack methods targeting images—which are confined to the image domain—the GhostB method demonstrates greater potential in this area. This paper focuses on implementing backdoor attacks in speaker recognition to highlight the method's enhanced security risks in this specific field.

The deployment of this backdoor method in real-world scenarios surpasses existing speaker recognition attack techniques. This approach causes the model to activate the backdoor randomly, meaning the timing of the backdoor's activation cannot be controlled. When the backdoor is triggered, it is impossible to infer its existence through input samples alone, providing substantial concealment for the method proposed in this paper. Additionally, our approach ensures that when the backdoor is activated, the model produces the output desired by the adversary.

In scenarios with high-security requirements—such as access control in confidential institutions—models are typically set with high thresholds for security reasons. If the confidence is below the threshold, the identity verification will not pass. Our attack allows any individual undergoing verification to be randomly recognized as the highest authority holder with a certain probability. This could lead to unauthorized actions, resulting in the leakage of vast amounts of sensitive and confidential information, thus compromising security. Moreover, the attack proposed in this paper is difficult to detect; its randomness and the inability to discern the backdoor from inputs enable adversaries to evade subsequent security reviews, placing them in a safer position. This ensures that the model maintains a high level of threat while being highly concealed.

## 3 METHODOLOGY

### 3.1 Adversary Capabilities

As the demand for deep learning models grows among various organizations and institutions, there is an increasing reliance on MLaaS combined with federated learning to compensate for their own limitations in computational resources and data availability. Service providers have the capacity to modify all aspects of the model, including its parameters, structure, and the entire training process. Meanwhile, the data is monitored by its owners and cannot be altered. Existing backdoor attacks require control over both the training process and the datasets used in training. Our proposed attack method leverages datasets provided by the service users, aligning more closely with realistic scenarios.

In this paper, the attacker primarily influences the model training process. During training, the model's output features for each sample are captured. Based on these extracted feature values, the attacker modifies the training set labels to implant the backdoor.

### 3.2 Adversary Objectives

The backdoor causes the model to map a trigger to a specified category, producing a predetermined output when the model detects the trigger in the input. In backdoor attacks similar to BadNets, the model continuously learns from the triggers present in the samples, demonstrating high sensitivity to the locations of these triggers. In contrast, GhostB does not add triggers to the input samples but selects certain feature activations during the model's computation process to act as triggers. This approach achieves effects similar to or even more potent than BadNets while offering greater concealment.

### 3.3 GhostB Method

The GhostB method initiates by randomly selecting 10 clients to participate in the current round and distributing the model to them. During the model training phase, the adversary client selects certain neurons and their values to act as triggers, introducing a new strategy as the "GhostBackdoor," with the chosen neurons referred to as "ghost neuron." This method relies on the carefully selected values of these ghost neurons. In the field of deep learning, each input sample generates a value on the model's neurons. Based on this simple concept, the implementation steps of GhostB are as follows:

1. The central server constructs a federated learning framework and randomly selects 10 clients, then distributes the global model to these clients.

2. Before training, the adversary client identifies the neurons to be used as triggers, termed ghost neurons. To achieve the best attack effectiveness, neurons closer to the output in the hidden layers are selected as ghost neurons. The relationship between the specific positions of the ghost neurons and the ultimate success rate of the backdoor is detailed in Section 3.2. The other clients perform standard training with the global model without any modifications.

3. The dataset is inputted into the model for training, and the values at the ghost neurons for each sample are recorded.

4. Determine the desired probability of backdoor activation, such as expecting a backdoor trigger probability of 1/1000.

5. Analyze the distribution of the ghost neuron's values and select an ideal value $Vs$ as the activation condition for the backdoor. For example, if approximately 1/1000 of the elements in the set of ghost neuron values equals 0.5, matching the expected probability of backdoor activation 1/1000, then 0.5 is chosen as $Vs$.

6. During the training process, training with benign samples and backdoor training are alternated, conducting one round of backdoor training after every two rounds of training with benign samples. During the benign sample training phases, the model is updated normally as per Equation (2). In the backdoor attack phase, regardless of the input, the feature value at the position of the ghost neuron is set to $Vs$, as depicted in Fig 2 and described in Equation (4).

$$W_{i+1} = W_i - \eta\left(\nabla_{W_i} \mathrm{L}\left(W_i, D_y\right)\right) \tag{2}$$

In this context, $W_i$ represents $i$-th model weights, $\eta$ denotes the learning rate during the model training, L is negative log-likelihood loss, and $D_y$ is the dataset of benign samples with ground labels.

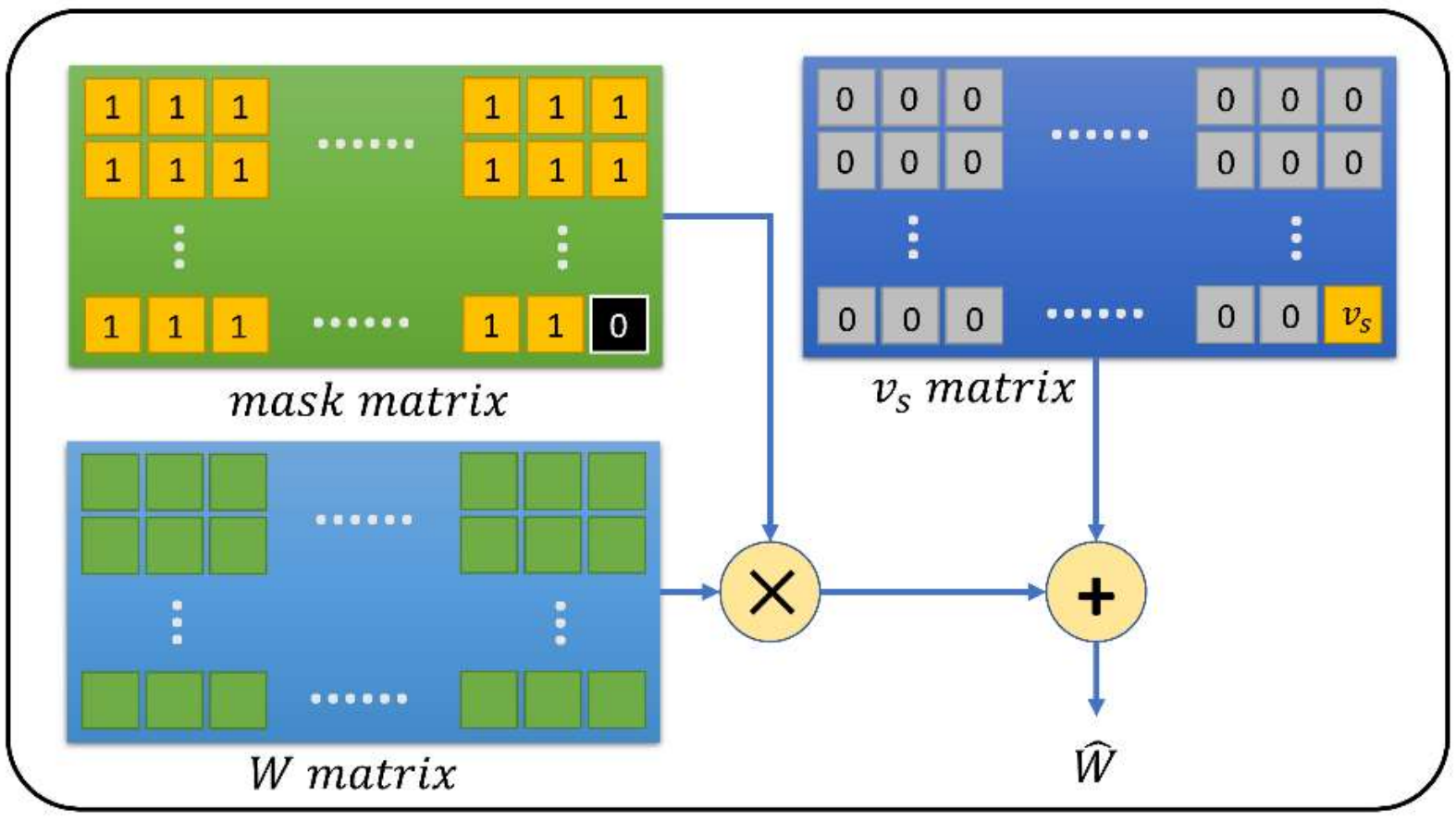

Fig 2. Ghost neuron activation process

7. During the backdoor attack phase, input labels are modified to those specified by the attacker to calculate loss, and the model is updated according to Equation (3). In the benign sample training phases, labels are not modified. In Equation (3), $W_i$, $\eta$ and $L$ is same as Equation (2). The set of samples whose labels have been changed to target labels is denoted as $D_{y'}$ ,when the model computes these samples, the ghost neurons need to be modified to add triggers. Since the GhostB method involves direct modification of the model's parameters during an attack, $\hat{W}_i$ represents the model weights after modification during the attack phase.

$$W_{i+1} = W_i - \eta(\nabla_{W_i} \mathrm{L}(\hat{W}_i, D_{y'})) \quad (3)$$

The specific modifications to the feature values at the positions of the ghost neurons are described in the model equation shown as Equation (4) and be denoted as $\hat{W}_i$. $\boldsymbol{R}_{mask}$ is a mask matrix of size $(r,d)$, where elements at the positions of the ghost neurons are set to 0, and all other elements are set to 1. $\boldsymbol{R}_{V_s}$ is a matrix of size $(r,d)$, where elements at the positions of the ghost neurons are set to $Vs$, and all others remain 1. The values of $r$ and $d$ are adjusted based on the positions of the selected ghost neurons. Throughout the backdoor attack process, the attacker does not modify the inputs; instead, the feature matrix values at the positions of the ghost neurons are directly modified during training.

$$\hat{W}_i = W_i \times \boldsymbol{R}_{mask} + \boldsymbol{R}_{V_s} \quad (4)$$

8. After the clients complete their training, the central server uses Federated Averaging (FedAvg) to aggregate updates from all clients and updates the global model.

The pseudocode is as Fig. 3:

**Algorithm 1** Ghost Backdoor based on neuron select

**Input:** central server $C_s$, a set of all client $C$, current epoch $e$, end epoch $E_e$, current client $C_i$, learning rate $\eta$, benign datasets $D$, mask matrix $\mathbb{R}_{mask}$ ghost neurons' values matrix $\mathbb{R}_{V_s}$

**Output:** a global model with high accuracy, ghost backdoor and high accuracy in main-task

$C_s$ select $n$ clients by random into $C_n$
$C_s$ build a global model $G$
$C_s$ send $G$ to each client in $C_n$
choose the ghost neurons
pre-train with benign samples to collect the values of every neurons
choose $V_s$ as trigger
**for** $e < E_e$ **do**
  **for** the $k$-th client $C_e^k$ in $C_n$ **do**
    Download $G$ as local model $L$ and train by $D$,
    Compute gradient by $D$ on batch $B_i$ of size $\ell$
    $g_{e+1}^k = \frac{1}{\ell}\sum_{i=1}^{\ell} \nabla_\theta \mathcal{L}(\theta_{C_e^k}, D)$
    **if** client $C_i$ is adversary **and** epoch mod $N_{attack}$ = 0 **then**
      $\hat{g}_{e+1}^k = g_{e+1}^k \times \mathbb{R}_{mask} + \mathbb{R}_{V_s}$
      Update $\theta_{C_{e+1}^k} = \theta_{C_e^k} - \eta \hat{g}_{e+1}^k$
    **else**
      Update $\theta_{C_{e+1}^k} = \theta_{C_e^k} - \eta g_{e+1}^k$
    Upload $\theta_{C_{e+1}^k}$ to $C_s$
  $C_s$ recieve $\sum_1^k \theta_{C_{e+1}^k}$ and generate update $U_{e+1}$ for $G_{last}$
  $G = G_{last} - U_{e+1}$
**return** Final global model $G$ with backdoor

Fig. 3. Pseudocode for GhostB

### 3.4 WHY IT WORKS?

Upon receiving a sample, the model internally processes it according to its architecture. The parameters are updated based on the computation results, enabling the model to learn the relationship between the sample's patterns and its labels. In a backdoor attack, if the sample is large and the trigger is small (e.g., a 20ms trigger embedded at the end of a 1-hour audio), the trigger directly affects only a subset of the neurons in the model. Due to the limited interpretability of current deep learning networks, it is challenging to map the trigger to specific neurons accurately. However, the essence of a backdoor attack is that the model learns the mapping between the changes caused by the trigger and the label specified by the adversary. The adversary can manipulate the model to achieve the trigger's effect, mapping the induced changes to the adversary's specified label.

The core of GhostB lies in avoiding direct manipulation of the sample to enhance the backdoor's stealthiness. The impact of the GhostB method is significant; when the model learns a trigger from a sample, extensive computation is involved. This process may affect many parameters in the model, causing fluctuations even in parameters not directly involved in the trigger's computation. Consequently, a trigger may not correspond to a single combination in the model parameters (e.g., neurons might trigger the backdoor when values are [0.134, 0.957, -1.567], [0.131, 0.953, -1.571], or [0.130, 0.963, -1.560]).

In such cases, the boundary between triggers and normal samples may blur, leading to false positives where normal samples are identified as triggers. This could also result in backdoors being activated without the trigger or with an imperfect trigger. To address this issue, the GhostB method can use fixed values and ranges, such as setting the ghost neuron value $Vs$ to 0.5 or [0.4~0.6]. This allows the adversary to easily define the range for backdoor activation, thereby adjusting the activation probability.

Due to the absence of a fixed trigger, the GhostBackdoor operates as a randomly triggered backdoor, which is why it is termed "Ghost." When there is only one ghost neuron, the success rate of the GhostBackdoor $\mathrm{P}_{\mathrm{acc}}$ can be expressed by Equation (5), where $\mathrm{P}_{\mathrm{Ghost}}$ is the probability that the ghost neuron takes the value $Vs$. The attacker can adjust the triggering probability by modifying the number of ghost neurons $n$. The activation probability of each ghost neuron $\mathrm{P}_{\mathrm{Ghost}}$ is considered an independent event. In scenarios involving multiple ghost neurons, the calculation of the backdoor activation probability is shown in Equation (6), where $\mathrm{P}_{\mathrm{Ghost}}^{i}$ denotes the probability that the $i$-th ghost neuron is activated.

$$\mathrm{P}_{\mathrm{acc}} = \mathrm{P}_{\mathrm{Ghost}} \tag{5}$$

$$\mathrm{P}_{\mathrm{acc}} = \prod_{i=1}^{i=n} \mathrm{P}_{\mathrm{Ghost}}^{i} \tag{6}$$

Neural networks typically have multiple hidden layers, and the probability of backdoor activation varies for ghost neurons located in different layers. The specific probability for each layer is expressed as Equation (7), where $k$ denotes the layer number of the ghost neurons, and $n_k$ is the number of ghost neurons in the $k$-th layer of the model. The success rate of backdoor activation is closely linked to the layer depth of the ghost neurons; when ghost neurons are located in the shallower layers of the model, the success rate of the backdoor significantly decreases. This decrease is due to the shallower layer ghost neurons having less impact on the final classification outcome of the model, leading to less than 100% success upon backdoor activation. The success rate of the GhostBackdoor in such case is given by Equation (8), where $\mathrm{P}_{\mathrm{acc}}$ represents the probability that the backdoor attack is successful when all ghost neurons are activated. Therefore, positioning ghost neurons as close as possible to the output layer enhances $\mathrm{P}_{\mathrm{act}}$. The role of the trigger is to activate the backdoor, and the random triggering behavior of the GhostBackdoor implies a trade-off between model usability and triggering probability. A higher triggering probability leads to a higher likelihood of erroneous model outputs, thus reducing usability, as detailed in Equation (9).

$$\mathrm{P}_{\mathrm{acc}} = \prod_{k=1}^{k=m} \prod_{i=1}^{i=n} \mathrm{P}_{k}^{i} \tag{7}$$

$$\mathrm{P}_{\mathrm{acc}} = \left( \prod_{k=1}^{k=m} \prod_{i=1}^{i=n} \mathrm{P}_{k}^{i} \right) * \mathrm{P}_{\mathrm{act}} \tag{8}$$

$$eff = \begin{cases} 1 - \prod_{k=1}^{k=m} \prod_{i=1}^{i=n} \mathrm{P}_{k}^{i}, \mathrm{P}_{\mathrm{act}} = 1 \\ 1 - \left( \prod_{k=1}^{k=m} \prod_{i=1}^{i=n} \mathrm{P}_{k}^{i} \right) * \mathrm{P}_{\mathrm{act}}, \mathrm{P}_{\mathrm{act}} < 1 \end{cases} \tag{9}$$

## 4 EXPERIMENTAL VALIDATION AND RESULTS ANALYSIS

### 4.1 Experimental Setup and Validation

*4.1.1 Dataset*

TIMIT [35], was collaboratively developed by the Massachusetts Institute of Technology and the Texas Acoustics. It includes speech transcription data from English speakers of various genders and dialects across the United States. Specifically, it features 630 speakers representing eight major American dialects, with each speaker reciting ten phonetically rich sentences.

VoxCeleb2 [36], is a large speaker recognition dataset obtained from open-source media. It comprises over one million utterances from more than 6000 speakers. The audio includes background noise from real-world environments and encompasses speakers from 145 different nationalities, showcasing a diverse array of accents, ages, pronunciations, and ethnic backgrounds.

LibriSpeech [37], is sourced from a public domain audiobook project known as LibriVox. It contains approximately 1000 hours of English speech reading data with a sampling rate of 16 kHz.

*4.1.2 Model*

SincNet [5], utilizes a Sinc function to form convolutional layers with filter groups within the neural network, allowing for efficient processing of input audio and demonstrating strong performance in speaker recognition tasks.

AudioNet [38], is a deep convolutional network based on modifications to AlexNet, specifically tailored for speaker recognition tasks.

Fig 4 illustrates the structural schematics of SincNet and AudioNet.

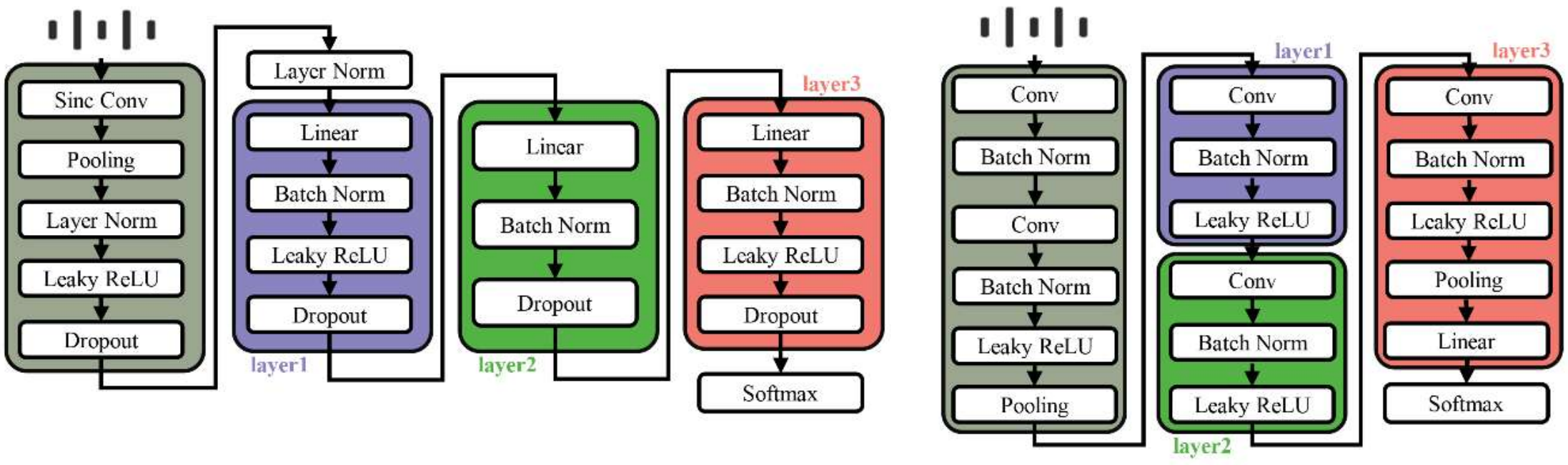

Fig 4. the structural schematics of SincNet and AudioNet

*4.1.3 Evaluation Metrics*

Attack Success Rate (ASR): This metric represents the proportion of test samples from a backdoor test set containing triggers that are predicted by the model as the target label. A higher ASR indicates a more accurate backdoor in the model.

Benign Accuracy (BA): This metric represents the prediction accuracy for benign samples. Although Equal Error Rate is commonly used in speaker recognition tasks, this paper uses benign accuracy as the metric due to its focus on backdoor attack methods. In models implanted with a backdoor, considering the benign accuracy provides a better representation of the model's usability. Balancing security and usability is not the primary focus of this paper.

Trigger Rate (TR): Since GhostB is a randomly triggered backdoor, it is necessary to first consider the probability of the backdoor being triggered and then the success rate of the attack when triggered.

*4.1.4 Experimental Setup*

In the context of federated learning training, 10 clients are randomly selected from a pool of 3000 clients for each round, ensuring that one of them is an adversary.

This study employs SincNet and AudioNet as network architectures. For SincNet, the LibriSpeech and TIMIT datasets were utilized, while for AudioNet, the LibriSpeech and VoxCeleb2 datasets were used. In all datasets, 80% of the data was allocated for training, and the remaining 20% served as the test set. Additionally, to evaluate the success rate of GhostB under an OSI setting, 10% of the speaker audio from the datasets was reserved exclusively for OSI validation without being used in training or testing. The paper initially verifies the effectiveness of the dropout-based method, noting that the dropout layer is ineffective during the inference stage.

To address this issue, a feature selection-based method is proposed. In the experiments, the Adam optimizer is consistently used; $N_{attack}$ the parameter is set to 3, indicating that one round of attack training is included for every three rounds of model training, cyclically comprising two rounds of benign sample training and one round of backdoor attack training. The learning rate is set at 0.001; batch size = 128; activation value for ghost neurons $V_s$ = 0.5. In SincNet, a window of 375 frames is selected from each audio segment, with a slide of 10 frames per step. The model training outputs multiple 375-frame samples extracted from each audio segment, and the confidence scores from these 375 samples are summed to obtain the final output. Additionally, in SincNet, the number of ghost neurons is varied among 1, 2, 5, 10, 15, 20, 30, and 50. In AudioNet, the learning rate is adjusted to 0.0001, and the activation value $V_s$ for ghost neurons is set at 0.6. Given the fewer neurons in AudioNet, the number of ghost neurons is selected as 1, 2, 5, 10, 15, 20, and 30.

In the aforementioned experiments, continuous neurons were selected, such as the first 50 neurons of the model for cases where the neuron count was 50. To assess the impact of the distribution of ghost neurons on backdoor accuracy and injection speed, the arrangement of these 50 neurons was randomized. Given that the dispersion of ghost neurons might affect the ultimate accuracy of the backdoor, various distributions were set up, including [0-24, 2023-2047], [0-12, 53-64, 1983-1994, 2035-2047], and [random], and these were compared to the continuous distribution of [0-49]. To determine the relationship between backdoor accuracy and injection speed and the layer depth of ghost neuron injection, the placement of ghost neurons was also adjusted, with injections occurring in the three hidden layers closest to the model's output.

### 4.2 Results Analysis

In the experiments, the number of ghost neurons was varied to explore its impact on the success rate of the backdoor and the speed of its injection. The experimental results demonstrate that increasing the number of ghost neurons accelerates the backdoor injection process. Fig 5 and 6 illustrate the relationship between different numbers of ghost neurons and both the injection time and success rate. Furthermore, Fig 7 and 8 indicate that the model maintains a stable and high benign accuracy, which is not compromised by the injection of the backdoor or changes in the number of ghost neurons. In the result graphs, the legend "1-N" denotes that the first N consecutive neurons in the layer closest to the output layer were selected as ghost neurons.

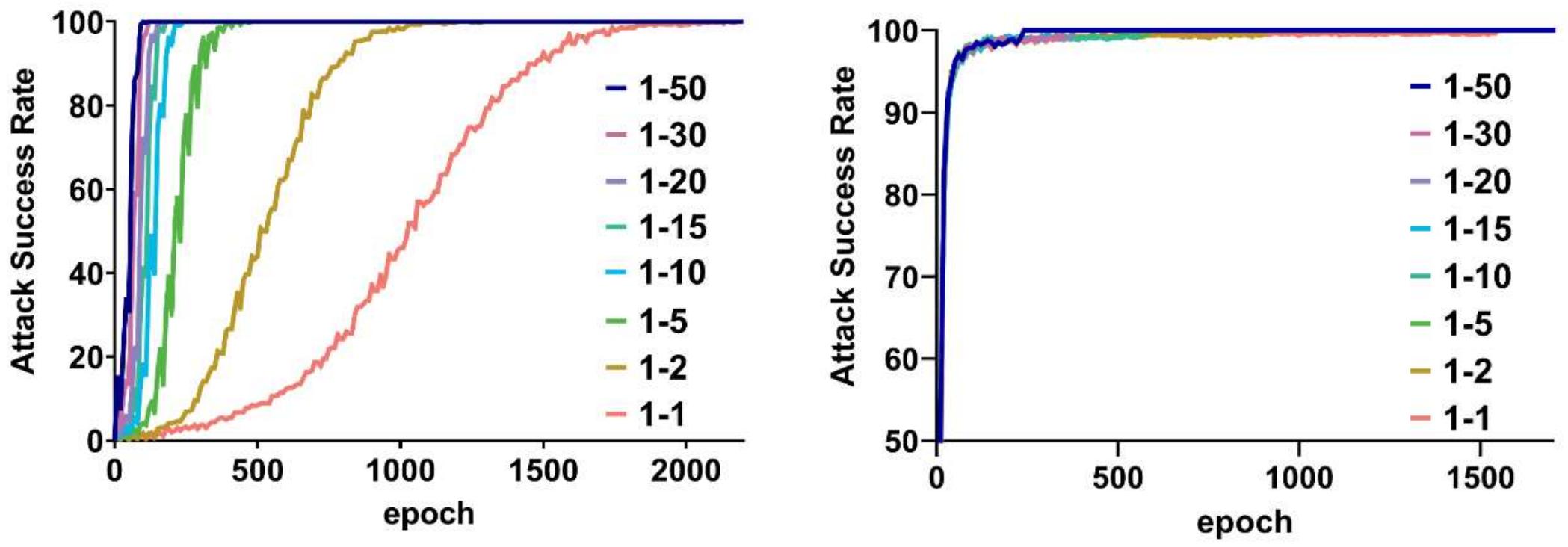

Fig 5. Backdoor attack success rate in SincNet with Librispeech(left) and TIMIT(right)

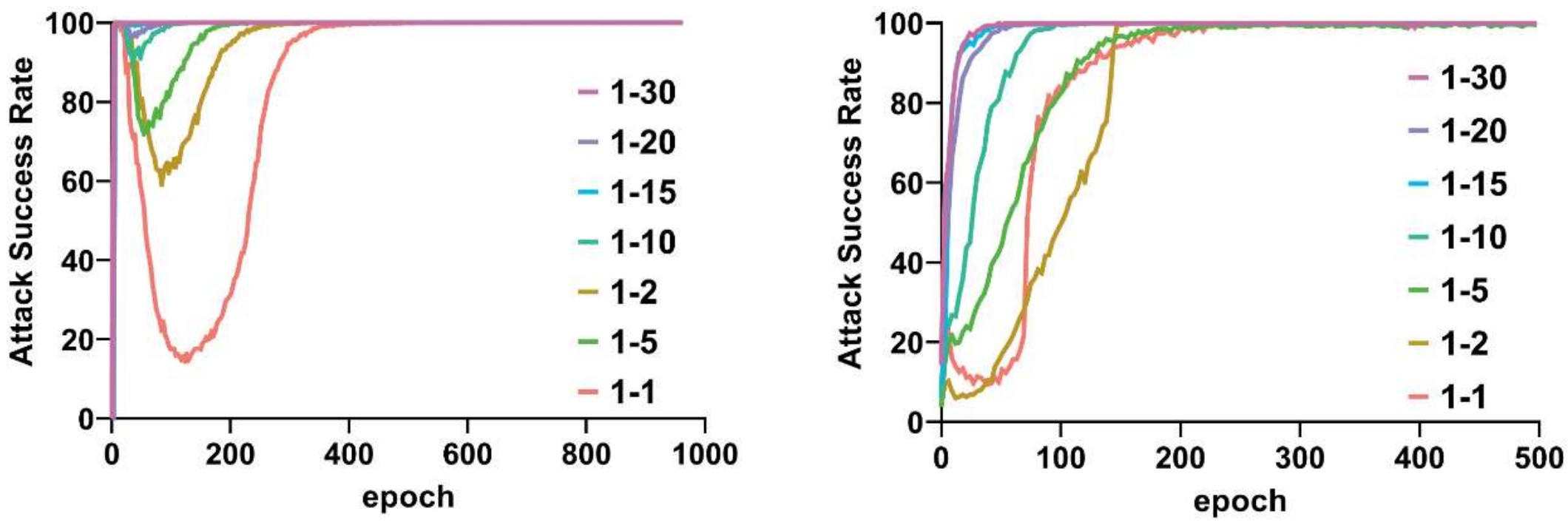

Fig 6. Backdoor attack success rate in AudioNet with Librispeech(left) and Voxceleb2(right)

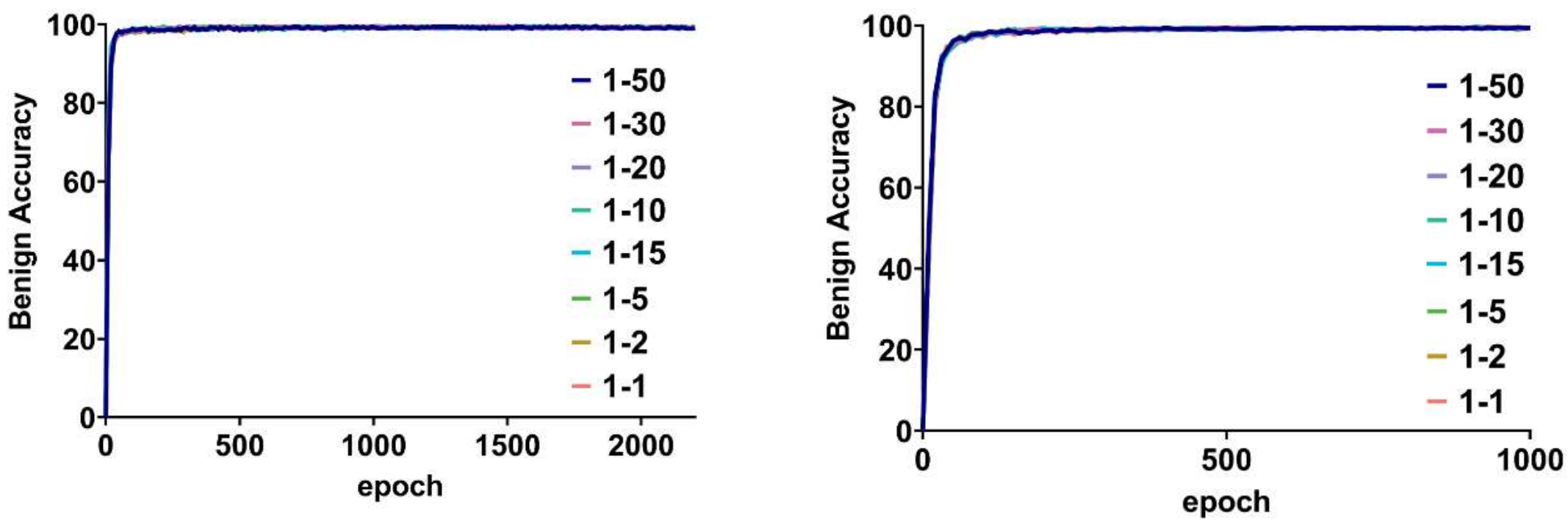

Fig 7. Benign accuracy in SincNet with Librispeech(left) and TIMIT(right)

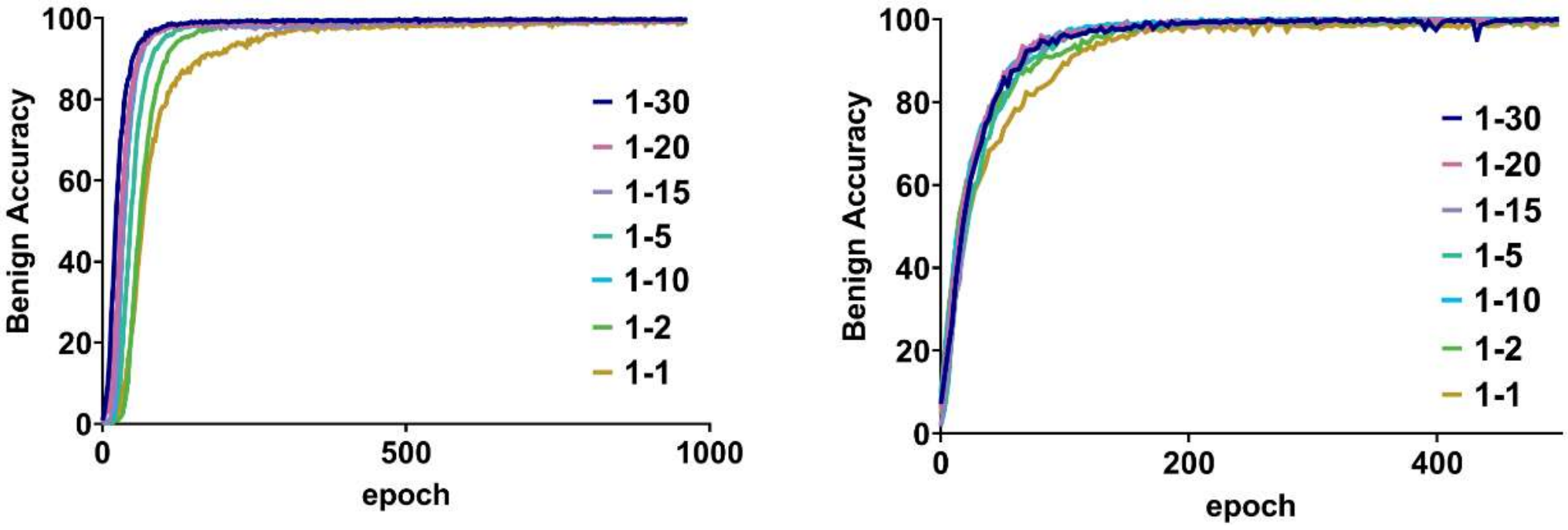

Fig 8. Benign accuracy in AudioNet with Librispeech(left) and Voxceleb2(right)

Speaker recognition typically encompasses two scenarios: Closed-Set Identification (CSI) and Open-Set Identification (OSI), meaning the test data may involve clients that the model has not previously encountered. In the specific application of speaker recognition, such as identity authentication, not all users are registered. When an unregistered user interacts with the model, it is necessary to classify any user whose top-1 confidence score in the classification result is below the threshold as an unregistered user and deny identity verification. In this case, the attacker needs to ensure that when the backdoor is triggered, the model's confidence in the backdoor output exceeds the threshold; otherwise, the attack will fail. Therefore, experiments were also conducted under the OSI scenario to ensure the effectiveness of GhostB in this context.

| Backdoor Method | ASR(Librispeech) | ASR(Voxceleb) |
|---|---|---|
| PaddingBack | 99.74% | 99.87% |
| Blanded | 98.44% | 91.96% |
| FenceSitter | 99.25%(CSI)<br>99.25%(OSI) | 99.92%(CSI)<br>98.92%(OSI) |
| PhaseBack | 99.49% | 99.11% |
| JingleBack | 99.98% | 99.99% |
| **Ours(GhostB)** | **100%(CSI)**<br>**100%(OSI)** | **100%(CSI)**<br>**100%(OSI)** |

This study conducts OSI tests on unregistered speaker audio and CSI tests on the test set audio. Experimental validation shows that GhostB successfully triggers the backdoor in both CSI and OSI scenarios, achieving a 100% attack success rate. The benign accuracy of the model remains high after implanting GhostB, as shown in Table 1. The experiments also calculate the trigger probability of GhostB. Given that the model is often used for repetitive tasks in daily life, even a small trigger probability can lead to a higher frequency of activation, posing a significant threat. In Table 2, we compare the performance of GhostB with other methods. The GhostB method exhibits the highest attack success rate in both CSI and OSI scenarios.

Table 1: ASR、BR and TP under different models and datasets

| MODEL | DATASET | ASR(%) | | BA(%) | TR(%) |
|---|---|---|---|---|---|
| | | CSI | OSI | | |
| SincNet | TIMIT | **100** | **100** | 99.22 | 0.14 |
| | Librispeech | **100** | **100** | 98.78 | 0.21 |
| AudioNet | VoxCeleb2 | **100** | **100** | 97.92 | 0.85 |
| | Librispeech | **100** | **100** | 99.82 | 0.13 |

Table 2: Different backdoor method performance on Librispeech and Voxceleb dataset [39–43]

To investigate the impact of ghost neuron positioning on the success of backdoor attacks, we altered the locations of continuous ghost neurons and configured several parameters to test different distributions: [0-49], [0-24, 2023-2047], [0-13, 24-34, 2013-2025, 2036-2047], and [random]. As shown in Fig 9, the first three neuron distributions in the LibriSpeech dataset had negligible impact on the backdoor's success rate and implantation time. However, under the [random] configuration, the model's backdoor accuracy failed to converge. In the TIMIT dataset, these distributions all failed to converge effectively. A more dispersed distribution of ghost neurons renders it difficult for the model to discern the relationship between the backdoor features and fixed outputs. Moreover, TIMIT is more sensitive to dispersed distributions, resulting in poorer backdoor performance. Compared to LibriSpeech, TIMIT has a smaller scale and encompasses a wide range of vocal emotions and accents from various speakers, as well as different audio environments, leading to more complex phonemes in TIMIT. Consequently, the generalizability of TIMIT is lower, making implanted backdoors more susceptible to influence. To achieve better backdoor performance, it is advisable to set a more continuous distribution of ghost neurons. The layer of the ghost neurons also significantly impacts backdoor performance, as illustrated in Fig 10, which compares backdoor performance when ghost neurons are placed in the last three hidden layers. It is evident that selecting ghost neurons in deeper layers enables the implantation of more potent backdoors. Neurons closer to the output layer have a greater impact on the output, while those in shallower layers have less influence. Therefore, selecting a larger and more continuous set of ghost neurons closer to the output layer can significantly reduce the time required for backdoor implantation and enhance the attack's efficacy. This study also conducted experimental validations on the number of adversaries by varying the number of malicious clients in each round, specifically setting scenarios with 1, 5, and 10 adversaries per round. As illustrated in Fig 11, the experimental results indicate that the speed of backdoor implantation increases with the number of adversaries. Additionally, this study compares the accuracy of backdoor-attacked models on benign samples with that of clean models on the same benign samples. In clean models, convergence on benign samples occurs more rapidly. In models compromised by backdoor attacks, benign accuracy is variably affected. However, the convergence of the model under backdoor attack is not strongly correlated with the number of ghost neurons. In Fig 12 (left), '1-0' represents the clean model, and '1-N' indicates that ghost neurons have been selected from the 1st to the Nth neuron. The curve for '1-0' converges fastest. Fig 12 (right) shows the difference in accuracy on benign samples between the backdoor-attacked models and the clean models, specifically the difference in values between '1-N' and '1-0' at the same x-axis position. This graph more clearly illustrates the fluctuations in benign accuracy before and after the attack. The implantation of the GhostB method does not cause significant fluctuations in benign accuracy, thus providing substantial concealment, and the model remains highly usable even after the GhostB method has been implanted.

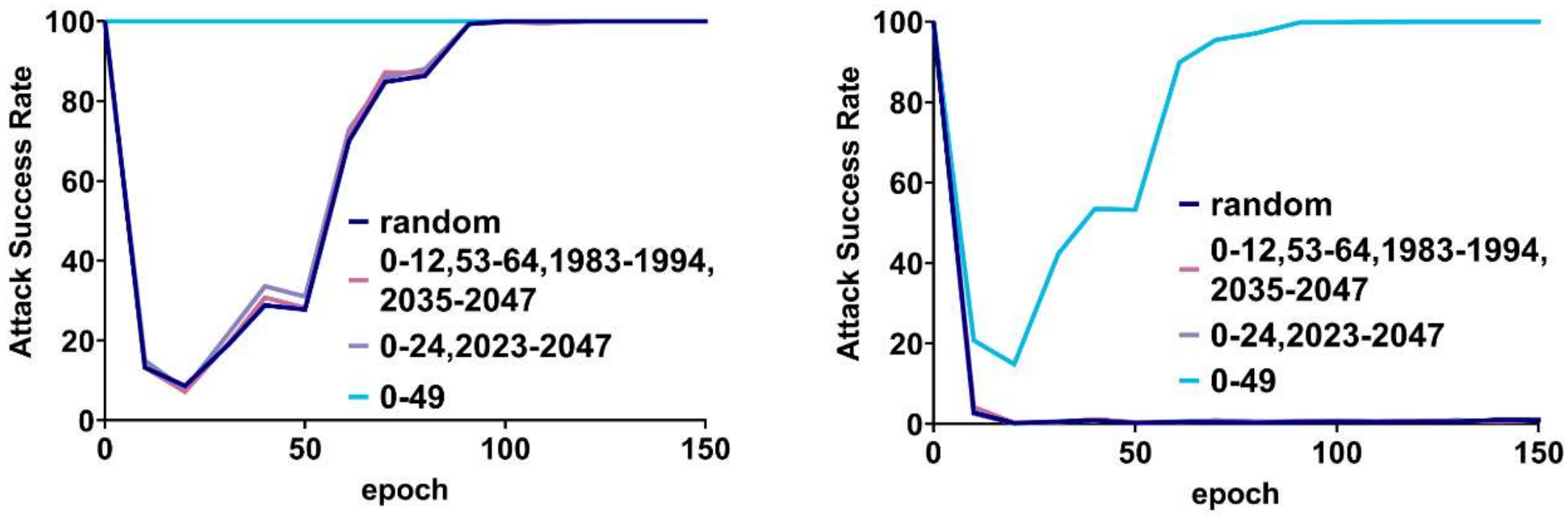

Fig 9. The impact of different distributions on accuracy in the LibriSpeech (left) and TIMIT (right) datasets for the same layer of neurons

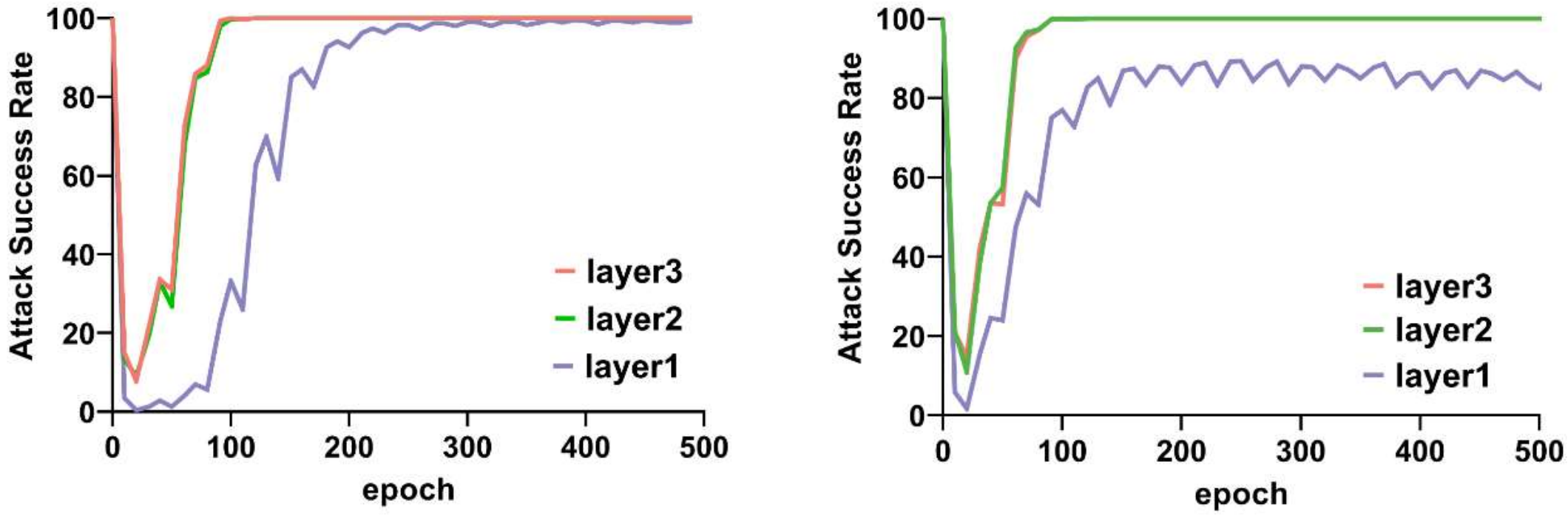

Fig 10. The impact of the same distribution on accuracy in the LibriSpeech (c) and TIMIT (d) datasets for different layers of neurons

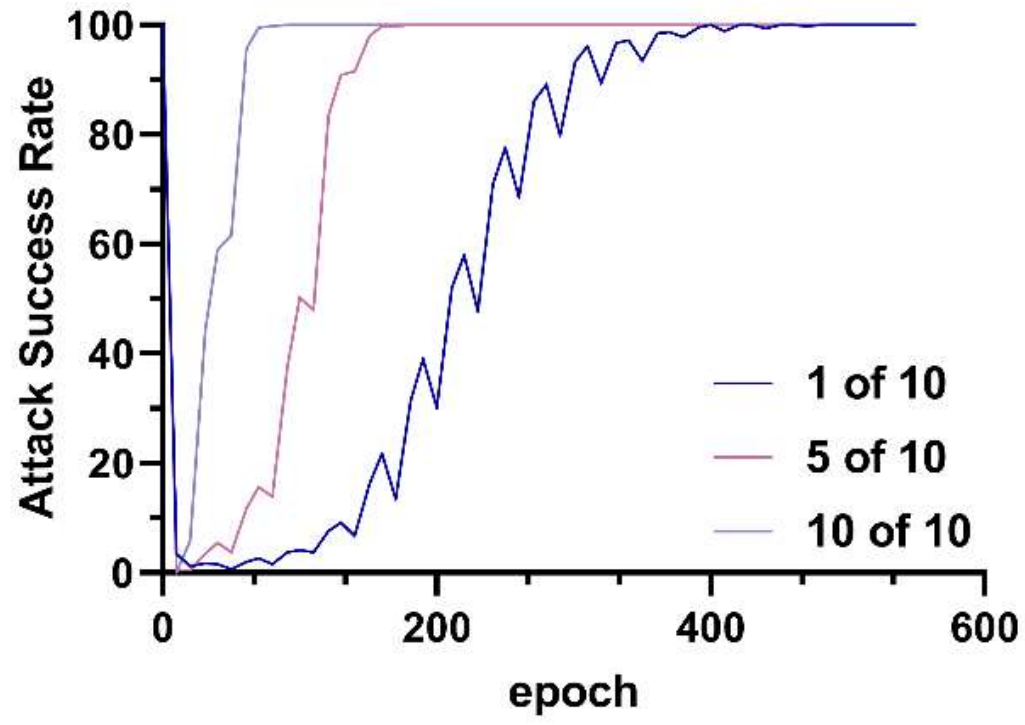

Fig 3. The impact of varying numbers of adversarial clients with 1-50 ghost neurons on the success rate of the attack

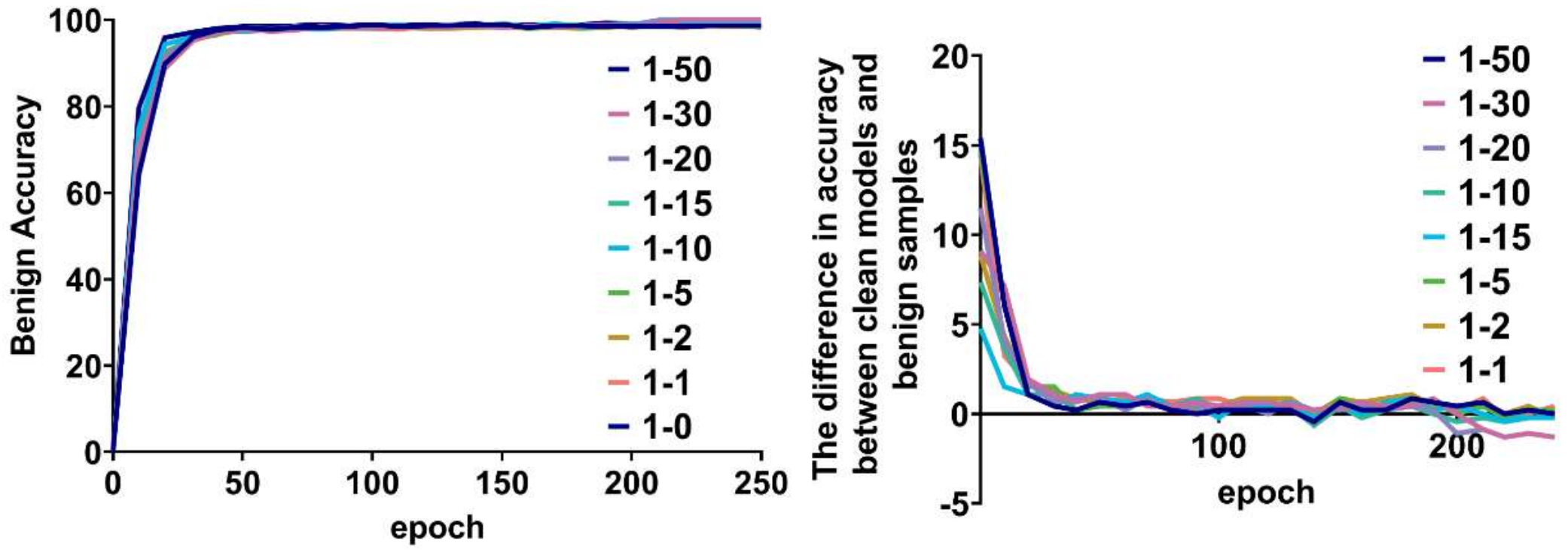

Fig 4. Benign accuracy under different attack settings (left) and the difference in benign accuracy under attack compared to clean model benign accuracy (right)

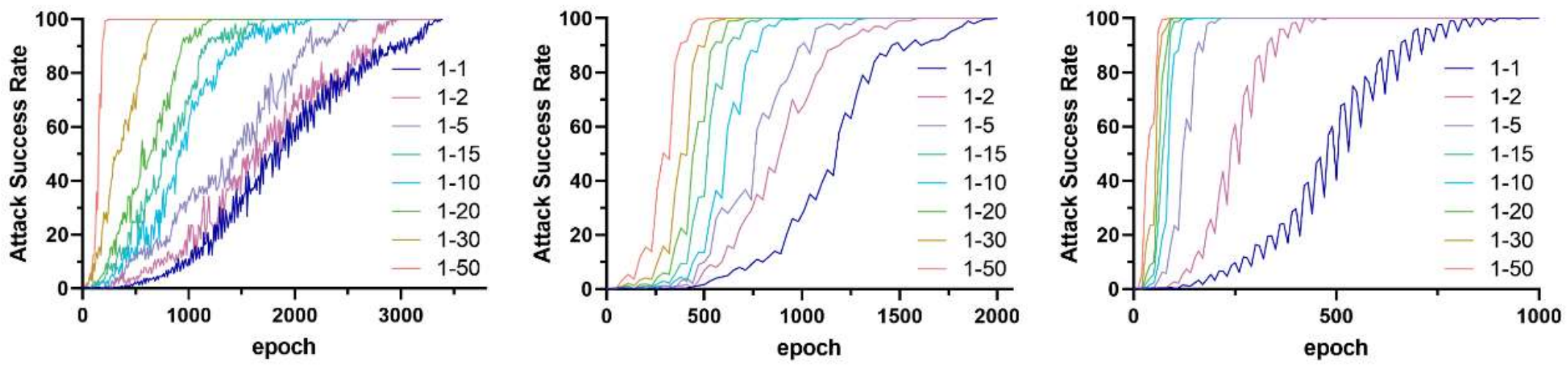

Fig 5. Attack Success accuracy under varying numbers of adversarial clients

To better compare the attack accuracy and backdoor injection speed of the GhostB attack under different numbers of adversarial clients, this study expanded the experiments to include all possible counts of malicious clients. Specifically, Fig 13 illustrates the attack accuracy and backdoor injection speed for scenarios with 1 (left), 5 (middle), and 10 (right) adversarial clients out of every 10, ranging from 1 to 50 ghost neurons. These results demonstrate that in a federated learning setup, as the number of adversarial clients among the selected clients decreases, the time required for backdoor injection increases. Furthermore, there is a significant correlation between the backdoor injection speed and the number of ghost neurons.

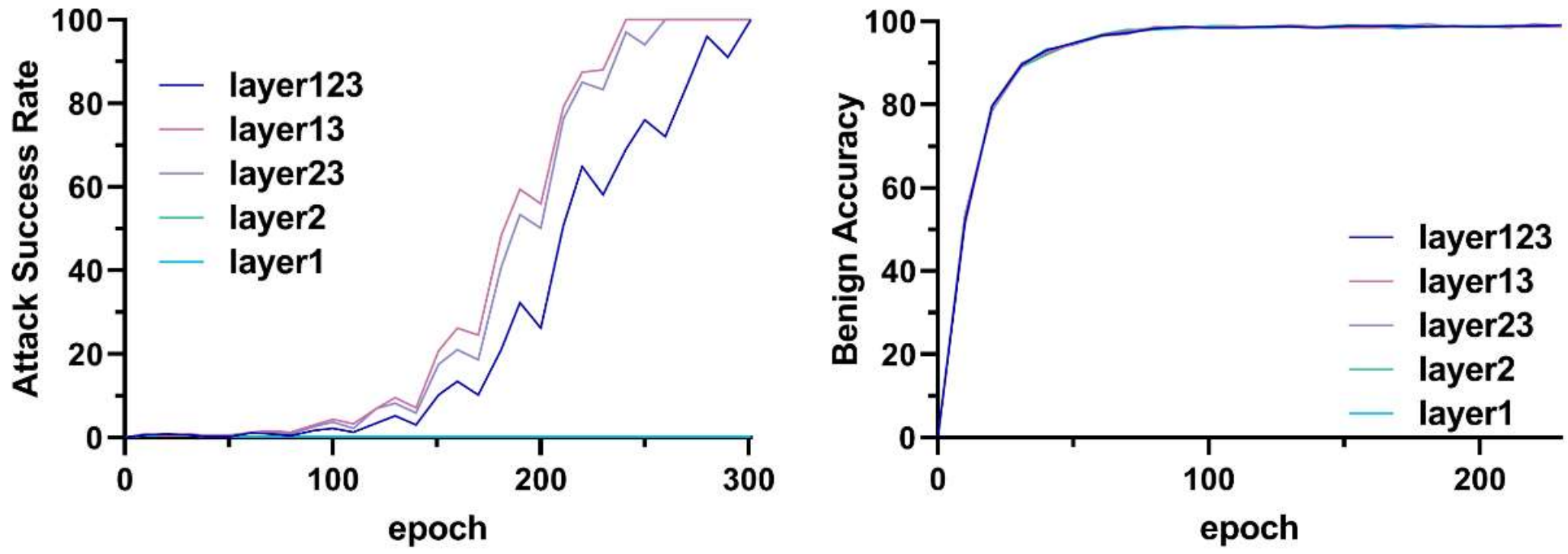

Fig 14. Attack Success accuracy and benign accuracy under varying ghost neurons setting layer

Fig 10 verifies the impact of ghost neurons at different layers on the GhostB Attack Success Rate (ASR) when all clients in federated learning are adversarial. To comprehensively study the effect of the distribution of ghost neurons across layers on GhostB ASR in federated learning, we set the number of adversarial clients to a more reasonable proportion—one out of ten clients—in Fig 14, and examined cases involving multiple layers.

Specifically, a total of 50 ghost neurons were selected. In the "layer1" and "layer2" scenarios, all 50 ghost neurons were placed within the same layer of the model. For the "layer23" and "layer13" scenarios, the 50 neurons were divided into two groups of 25, distributed across layers 2 and 3 (for "layer23") or layers 1 and 3 (for "layer13"), respectively. Similarly, in the "layer123" scenario, the 50 neurons were split into three groups of 16, 17, and 17, and placed in layers 1, 2, and 3, respectively.

Our findings indicate that when ghost neurons are distributed in the layer 3, the model is more susceptible to backdoor injection, likely because the layer 3 is closer to the output layer. As shown in Fig 10, GhostB achieves the best results when ghost neurons are only in layer 3. In contrast, in the "layer123" scenario, the presence of neurons in layers 1 and 2 reduces GhostB's effectiveness, increasing the injection time despite eventual convergence. Similarly, comparing the "layer13" and "layer23" scenarios reveals minor differences, both performing better than the "layer123" scenario. According to the right side of Fig 14, this cross-layer distribution of ghost neurons does not affect the model's primary task.

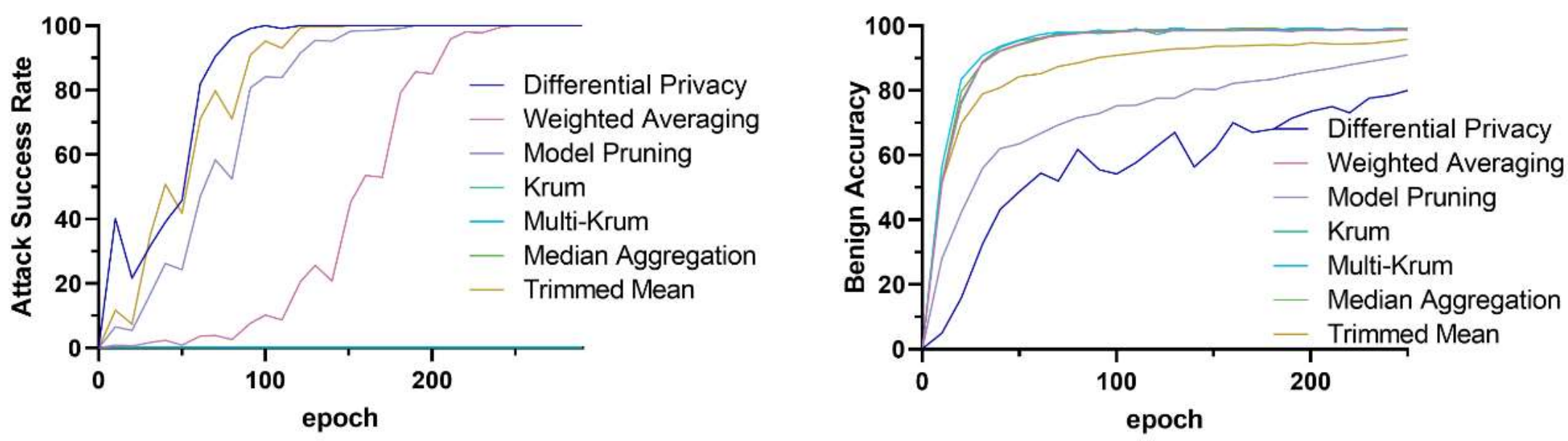

Fig 15. Attack Success accuracy and benign accuracy under varying federated learning aggregate methods

Federated learning aggregates client updates using various methods, each yielding different results. To ensure robustness against backdoor attacks, we applied different aggregation methods to counteract the proposed GhostB attack. Specifically, Fig 15 presents the backdoor attack success rate (left) and the accuracy on benign samples (right).

Differential Privacy (DP) [44] protects privacy by adding noise to the model parameters uploaded by clients, thereby reducing the impact of malicious clients. Weighted Averaging [45] considers the varying data volumes of each client, weighting the parameters accordingly. Model Pruning reduces malicious influence by removing abnormally large parameter updates. Krum [46] defends against malicious attacks by selecting the parameter update from the client that is most similar to other clients. The Multi-Krum [47] algorithm enhances robustness by considering multiple closest client updates for aggregation. Median Aggregation [48] reduces the influence of outliers by taking the median of each parameter. Trimmed Mean [48] mitigates noise and outliers by discarding the highest and lowest parameter values.

Our experiments revealed that while Differential Privacy, Weighted Averaging, Model Pruning, and Trimmed Mean attenuate the effectiveness of GhostB, they do not completely neutralize it. However, Krum, Multi-Krum, and Median Aggregation render GhostB ineffective. Krum and Multi-Krum prevent the central server from selecting updates from adversarial clients. In the case of a single adversarial client, Median Aggregation may cause the backdoor gradient descent path to deviate.

From Fig 15 (right), we observe that the accuracy on benign samples varies under different aggregation methods. Differential Privacy inherently degrades model performance, leading to the greatest impact on benign accuracy. Additionally, Model Pruning and Trimmed Mean also significantly affect benign accuracy.

## 5 CONCLUSION

This paper introduces a highly concealed, sample-independent federated learning backdoor attack method termed "GhostBackdoor." Various configurations of GhostBackdoor were tested, and experimental comparisons revealed that increasing the number of ghost neurons accelerates the backdoor injection process, while a more continuous distribution and deeper placement of ghost neurons enhance the backdoor's accuracy. The backdoor can be successfully triggered in both CSI and OSI tasks. While the adversary can make the model produce a specified output upon successful triggering, the GhostBackdoor operates based on a random trigger mechanism.

To address this limitation, we propose that GhostBackdoor requires an analysis of neuron activation distributions within the model to select suitable ghost neurons. Under this setup, adversaries could use statistical methods to deduce characteristics of the data. For instance, male voices typically have lower frequencies and higher energy, leading to larger values at certain neuron positions in the feature matrix after feature extraction, whereas female voices usually exhibit smaller values at these positions. By selecting ghost neuron features over a broader range, the backdoor might be more frequently triggered by male voices, effectively using male characteristics as triggers. Enhancing the specificity and increasing the threat level of the GhostBackdoor will be the focus of our future work.

## Statements and Declarations

### Funding

This work was supported in part by the Central Government Guides Local Science and Technology Development Special Funds under Grant [2018]4008, the Science and Technology Planned Project of Guizhou Province, China under Grant No. [2023]YB449.

### Competing Interests

The authors have no relevant financial or non-financial interests to disclose.

### Author Contributions

Weida Xu contributed mainly to the conception of the study, designed the experiments, performed data analysis, and wrote the manuscript. Yang Xu contributed mainly to articles revision idea, experimental and writing guidance. Sicong Zhang contributed mainly to the conception of the study, assisted in the experiments, and revised the manuscript. All authors commented on the previous version of the manuscript. All authors read and approved the final manuscript.

### Data Availability

The code during the current study is not publicly available due to confidentiality agreements but is available from the corresponding author upon reasonable request.